# Clustering Chinese Regional Cultures with Online-gaming Data

Xianwen Wang[a,b]*, Wenli Mao[a]

[a] WISE Lab, Dalian University of Technology, Dalian 116085, China.

[b] DUT- Drexel Joint Institute for the Study of Knowledge Visualization and Scientific Discovery, Dalian University of Technology, Dalian 116085, China.

* Corresponding author.

Email address: xianwenwang@dlut.edu.cn; xwang.dlut@gmail.com

**Abstract:** To identify cluster of societies is not easy in subject to the availability of data. In this study, from the prospective of computational social science, we propose a novel method to cluster Chinese regional cultures. Using millions of geotagged online-gaming data of Chinese internet users playing online card and board games with regional features, 336 Chinese cities are grouped into several main clusters. The geographic boundaries of clusters coincide with the boundaries of provincial regions. The north regions in China have more geographical proximity, when regional variations in south regions are more evident.

# Introduction

Unlike the massive data available for natural sciences, data in social science are hard to get (Latour, 2007). However, with the vast geotagged usage data generated by people in the web, it is possible for researchers to reveal and understanding details of both individual and social behavior with unprecedented detail (Eagle & Pentland, 2006; Girardin, Calabrese, Fiore, Ratti, & Blat, 2008; Lazer et al., 2009; Manovich, 2011), and create and define new methods of observing, recording, and analyzing human dynamics (O'Neill et al., 2006).

With the data collected from 100 Bluetooth-enabled mobile phone users for 9 months, Nathan Eagle & Pentland make a good try to capture daily human behavior, "identify socially significant locations, and model organizational rhythms" (Eagle & Pentland, 2006). Using the geotagged usage data of scientific paper downloading from Springer Verlag, Wang et al try to capture details of peoples' daily working behavior (Wang et al., 2013; Wang et al., 2012).

To identify cluster of societies is not easy in subject to the availability of data. There has been much effort to group countries into similar clusters using survey data (Brodbeck et al., 2000; Cattell, 1950; Gupta, Hanges, & Dorfman, 2002; Smith, Dugan, & Trompenaars, 1996) (Furnham, Kirkcaldy, & Lynn, 1994). Previous studies have shown that many factors, i.e. religious and linguistic commonality, geographic proximity, and mass migrations and ethnic social capital are relevant factors in the clustering of societies(Crang & Zhang, 2012; Gupta et al., 2002; Haandrikman & Hutter, 2012; Ho & Hatfield, 2011).

As one of the world's earliest civilization, most populous and second largest country by land area, China has diversified cultures. i.e. China has as many as 292 living languages, people in different cities usually can't standard each other if they speak dialect, especially in south China. For a long time, Chinese culture are roughly divided into two parts, south and north, along the Qinling Mountains-Huaihe River line. There are some other classifications, including three zones (the eastern, central, and western region), six economic zones (northeast, eastern, central and western economic region), etc. However, these classifications are rather rough and lack of evidence.

In this study, using a kind of novel large sample usage data, we try to cluster Chinese cultures from a new perspective.

# Data preparation

*QQ Game Map*

QQ is an abbreviation of Tencent QQ, which had been called OICQ during the period of 1998 – 2000. Now QQ is the most popular instant messenger software in China. By the end of 2012, there were 798 million active user accounts with approximately 170 million users online at a time.

QQ Game is a casual games client, offering only multi-player online games. 183 board and card games are available through the client. There are approximately 8 million active online players during peak hours, generally 21:00 – 23:00. Online players once peaked to 9.4 million in December 2012.

QQ Game Map is a service launched by Tencent, Inc. since October 22, 2012. It provides real-time visualization map of geographical distribution QQ game online players. Besides general distribution of the total QQ game players, the map also provide visualization for any single QQ game.

*Data*

Our data is collected from the website of QQ Game Map (http://qqgame.qq.com/online.shtml). For the 183 QQ games and 376 cities in China, the number of online players of each game in each city is recorded at the time 21:20, March 11, 2013, as table 1 shows.

**Table 1 Data collecting**

| City | Total players | Shanghai mahjong | Sichuan mahjong | … | Guangdong Jipinghu mahjong |
|---|---|---|---|---|---|
| Shanghai | 272,730 | 12,127 | 1521 | … | 103 |
| Beijing | 272,668 | 135 | 2011 | | 88 |
| Chongqing | 208,271 | 9 | 20,791 | … | 18 |
| Shenzhen | 161,071 | 22 | 1059 | … | 677 |
| … | … | … | … | … | … |
| All cities | 7,959,505 | 17,918 | 109,672 | … | 9848 |

# Results

*Geographical distribution*

Fig. 1 shows the geographical distribution of QQ game players at 21:00 on March 11, 2013. The node size is correlated to the number of players of the city. The white shining spot in the center of the node means that someone of the region is beginning to play game.

There are approximate 7.55 million players online at this time point. Most nodes are clustered together in the east coastal regions, north China plain (including Beijing, Tianjin and Shandong province), when the dark west China form a powerful contrast to the dazzling east regions.

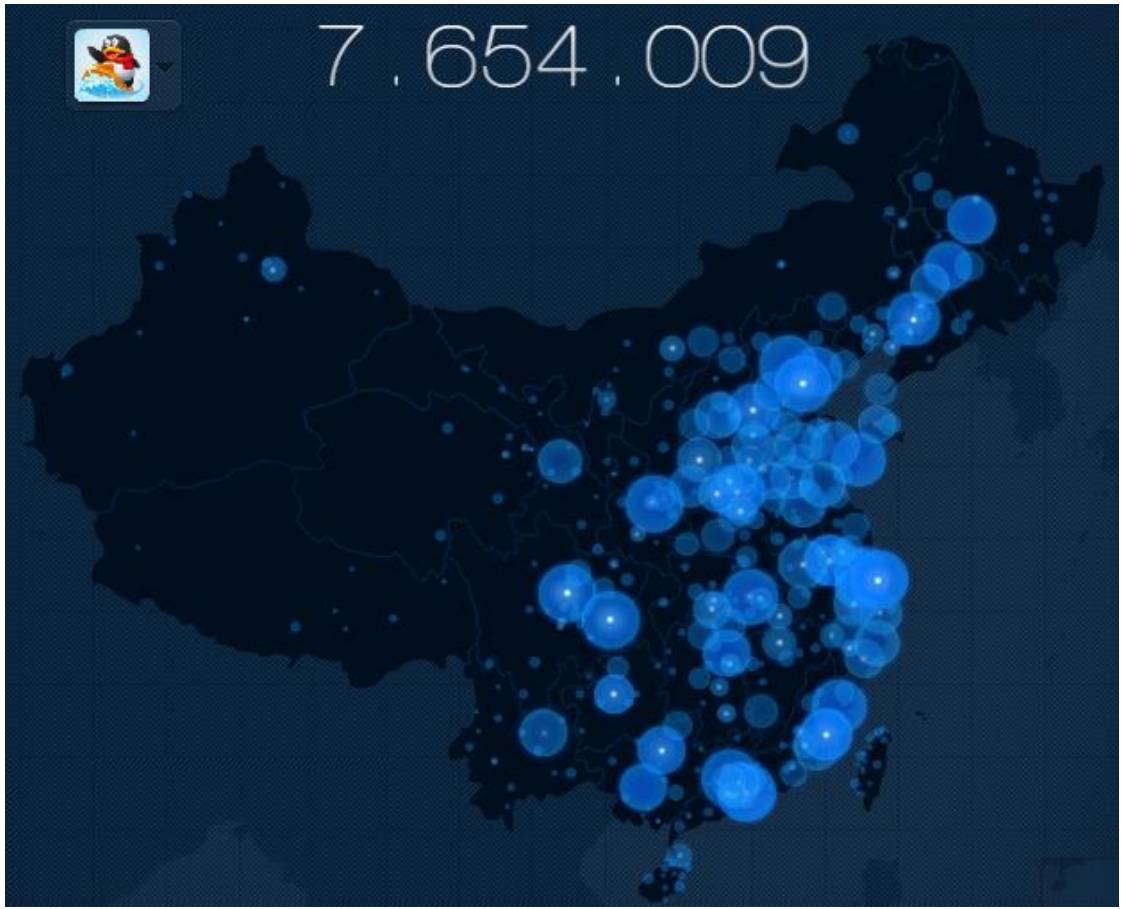

**Fig. 1.** Screenshot of QQ Game Map (21:00, March 11, 2013)

Shanghai is the most active city in China in terms of online game playing, with over 0.27 million players online at the time, which accounts for about 1.18% of its population. Beijing ranks number 2 with slightly less players as Shanghai but the second highest ratio of players to population (1.39%). The number 3 city is Chongqing with approximately 0.21 million players and a relative low player-to-population ratio (0.72%). It is worth mentioning that all the top 3 cities are municipalities directly under the China central government, which are usually big cities with large population.

With approximately 0.16 and 0.14 million online game users, Shenzhen and Guangzhou rank fourth and fifth correspondingly. Shenzhen is the youngest big city located in the south of China and is adjacent to Hong Kong, which owns a number of factories and armies of young migrant workers. It is not strange that Shenzhen has the highest player-to-population ratio because of the plenty of young people aged from 18 to 40. It is the same condition for Guangzhou, the capital city of Guangdong (Canton) province.

In Table 2, we list the top 20 cities with most QQ game players at the time of

21:00, March 11, 2013. The total number of the 20 cities is over 2.52 million and accounts for 31.82% of all QQ game players in the whole China.

Table 2 Top 20 cities with most QQ game players

| Rank | City | Province | Number of players | Population | Percent |
| --- | --- | --- | --- | --- | --- |
| 1 | Shanghai | NA | 272,730 | 23,019,148 | 1.18% |
| 2 | Beijing | NA | 272,668 | 19,612,368 | 1.39% |
| 3 | Chongqing | NA | 208,271 | 28,846,170 | 0.72% |
| 4 | Shenzhen | Guangdong | 161,071 | 10,357,938 | 1.56% |
| 5 | Guangzhou | Guangdong | 144,728 | 12,700,800 | 1.14% |
| 6 | Chengdu | Sichuan | 142,662 | 14,047,625 | 1.02% |
| 7 | Hangzhou | Zhejiang | 133,859 | 8,700,400 | 1.54% |
| 8 | Tianjin | NA | 123,559 | 12,938,224 | 0.95% |
| 9 | Wuhan | Hubei | 123,224 | 9,785,392 | 1.26% |
| 10 | Zhengzhou | Henan | 96,454 | 8,626,505 | 1.12% |
| 11 | Suzhou | Jiangsu | 94,578 | 10,465,994 | 0.90% |
| 12 | Shenyang | Liaoning | 93,993 | 8,106,171 | 1.16% |
| 13 | Xi'an | Shaanxi | 91,304 | 8,467,837 | 1.08% |
| 14 | Shijiazhuang | Hebei | 88,984 | 10,163,788 | 0.88% |
| 15 | Qingdao | Shandong | 83,830 | 8,715,100 | 0.96% |
| 16 | Dongguan | Guangdong | 83,826 | 8,220,237 | 1.02% |
| 17 | Ningbo | Zhejiang | 79,605 | 7,605,689 | 1.05% |
| 18 | Fuzhou | Fujian | 78,629 | 7,115,370 | 1.11% |
| 19 | Baoding | Hebei | 75,858 | 11,194,400 | 0.68% |
| 20 | Quanzhou | Fujian | 74,213 | 8,128,530 | 0.91% |
|  | All top 20 | NA | 2,524,046 | 236,817,686 | 1.07% |
|  | All China | NA | 7,932,716 | 1,339,724,852 | 0.59% |

Note: population data are collected from *Communiqué of the National Bureau of Statistics of People's Republic of China on Major Figures of the 2010 Population Census (No. 2)*

As shown in Table 1, among the top 20 cities, 3 are located in Guangdong province. Guangdong is the most populous province according to the 2010 population census of China. Due to the booming economy and high demand for labor, especially young labor, Guangdong has the massive influx of migrants from other provinces. And QQ game is very popular in Chinese young people. According to the statistics of

Tencent QQ, the number of QQ game players born in 1980 – 1989 accounts for 46.38%, as Fig. 2 shows.

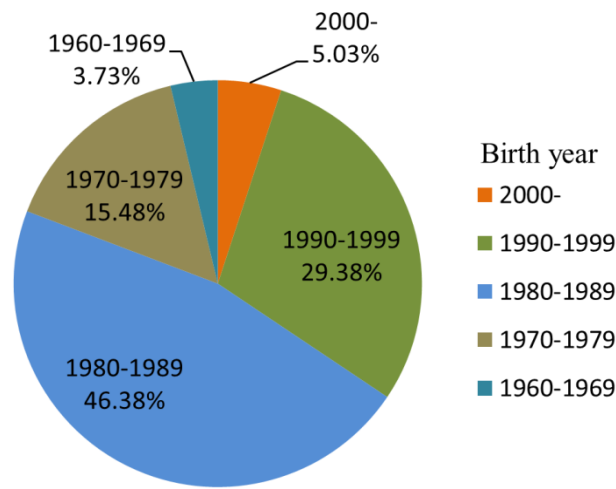

**Fig. 2.** Age groups of QQ game players

*Regional difference and cultural diversity in China*

Playing card and board game is a favorite pastime for Chinese people. China has strong board and card game culture. There are hundreds of card and board games in China. Most provinces and many cities have distinct card and board games. For example, there are more than 30 kinds of mahjong in China. Shanghai has Shanghai mahjong. Zhejiang, a neighbor province of Shanghai, has Hangzhou mahjong and Ningbo mahjong. So is the case for card games.

Fig. 3 shows the geographical distribution of Shanghai mahjong online players. The nodes are concentrated in the Yangtze River delta, including Shanghai, Jiangsu province and Zhejiang province. There is also scattering distribution in some other big cities, e.g. Beijing, Tianjin, Shenyang, Wuhan, Chengdu and Nanning.

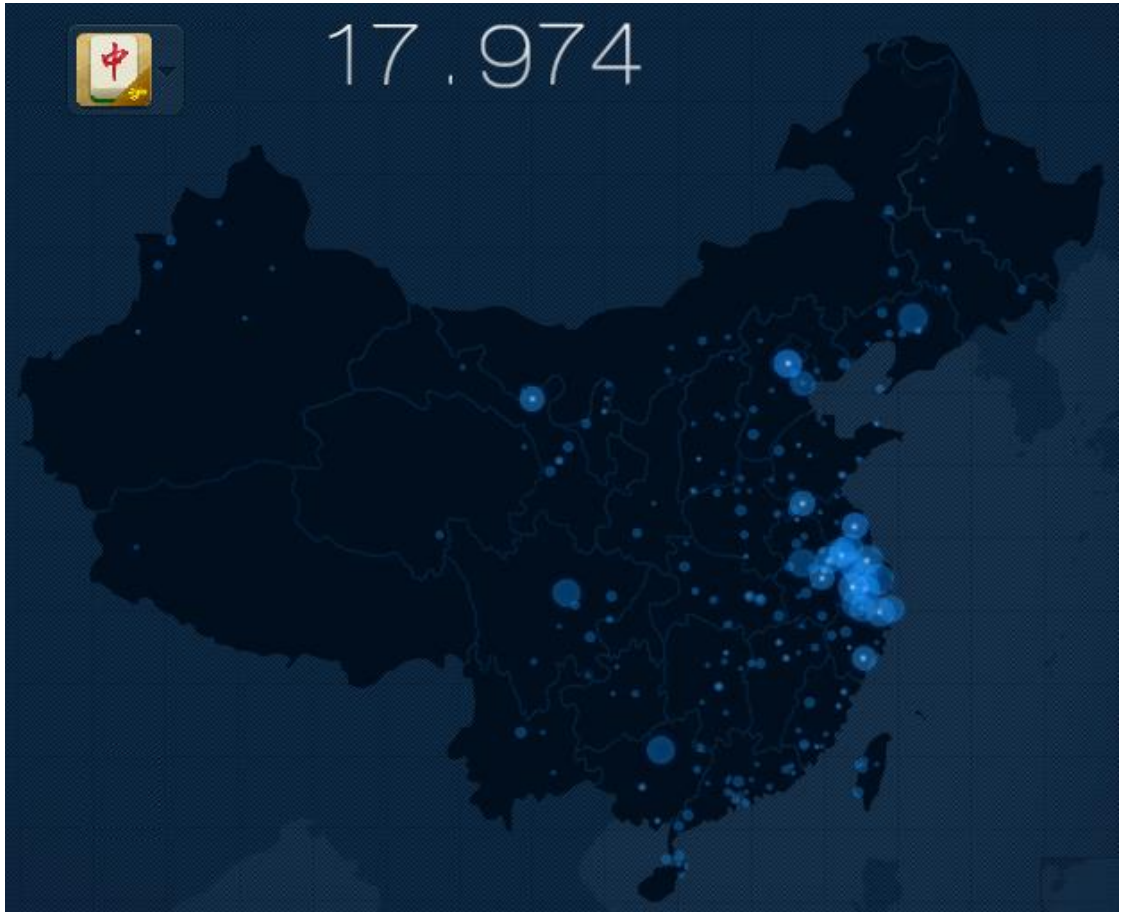

**Fig. 3.** Geographical distribution of Shanghai mahjong players

For another kind of mahjong game, Sichuan mahjong, as Fig. 4 shows, most players are concentrated in the west China, including Sichuan province, Chongqing city, Guizhou province, Kunming city. Other regions, e.g. Yangtze River delta in east China, Pearl River Delta in south China, also have scattered distribution.

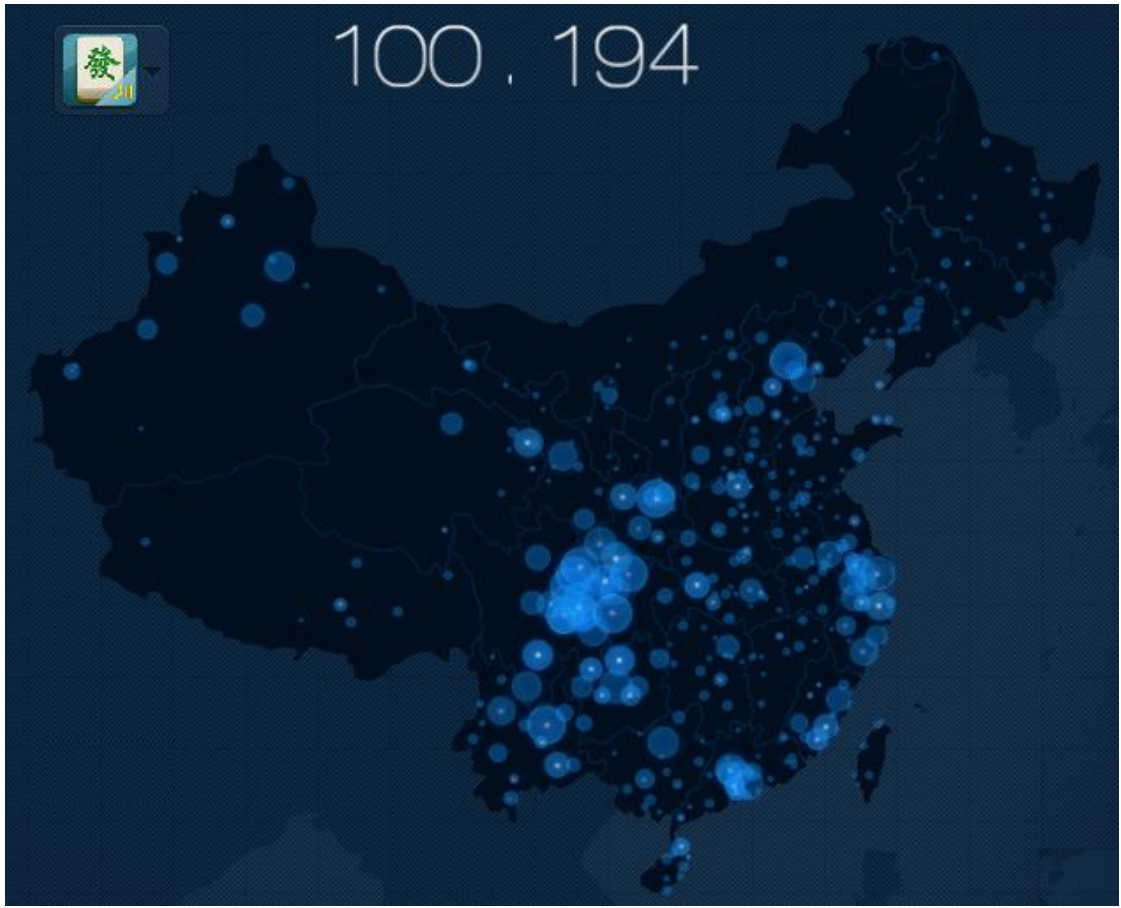

**Fig. 4.** Geographical distribution of Sichuan mahjong players

For contrast, we also choose another game as control sample, which is Chinese chess. It is a popular chess game in China and without much regional feature. As Fig. 5 shows, the nodes distribution is very consistent with the general distribution shown in Fig. 1.

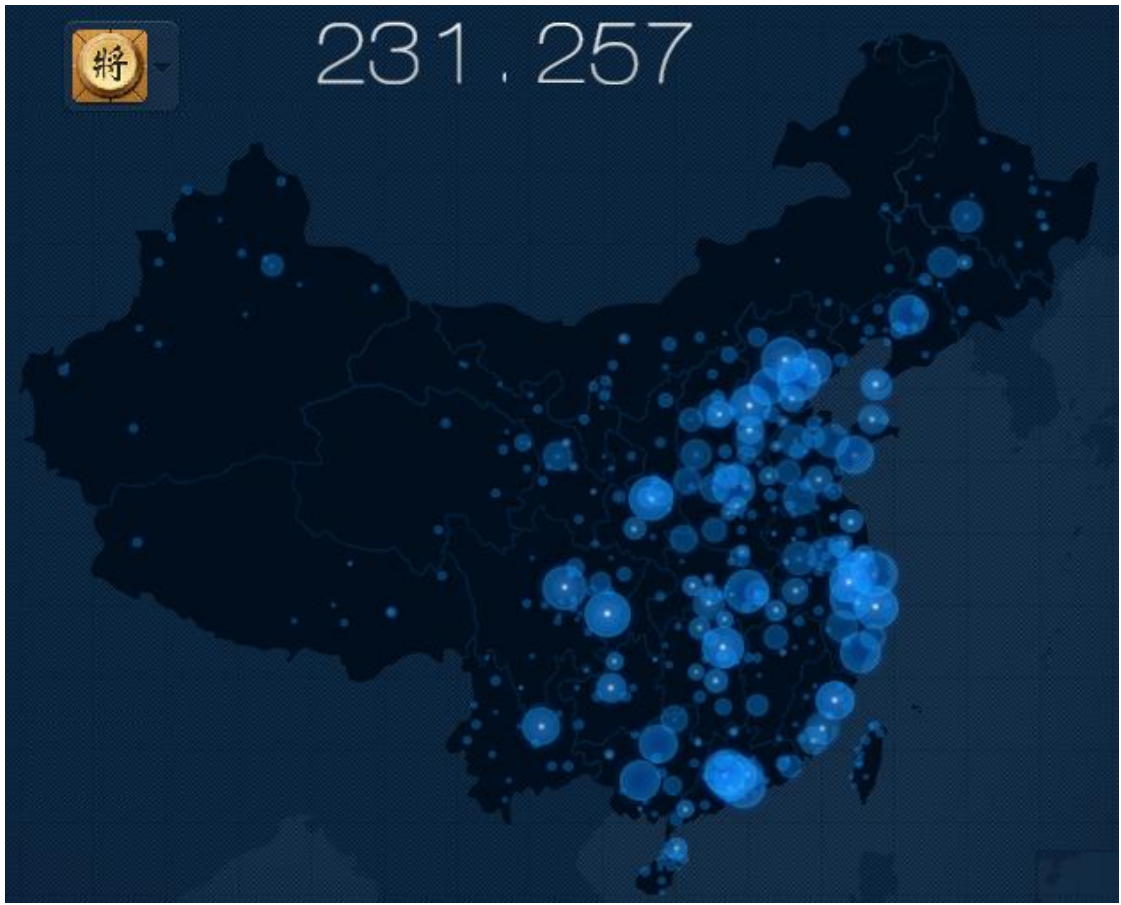

**Fig. 5.** Geographical distribution of Chinese Chess players

So, in order to identify the interregional similarities and interregional differences, only games with distinct regional features are included.

*Geographical Region Clustering*

For each provincial region in China, 2-3 kinds of QQ games with regional features are selected as research samples. Finally, 61 kinds of games in total are selected from the QQ game library.

For the 376 cities in China, we record the number of players of each game in each city. Considering the population size of cities, we divide the number of players of one game by the total players of all games in the city. Because of the data missing of some regions in Hainan province, and the special status of Taiwan, Hong Kong and Macau, these areas are excluded. Finally, 336 cities in Chinese mainland are selected as research objects.

To determine the optimal number of clusters which could best distinguishes feature similarities and differences, grouping effectiveness is measured using the

Calinski-Harabasz pseudo F-statistic (Caliński & Harabasz, 1974).

In ArcGIS 10.1, CONTIGUITY_EDGES_ONLY method is specified as spatial constraint to limit group membership to contiguous features, Minimum Spanning Tree algorithm is used for grouping (Assunção, Neves, Câmara, & Da Costa Freitas, 2006). For other parameter settings, EUCLIDEAN is selected as distant methods, and the number of neighbors is set as 8.

Fig. 6 shows the result of pseudo F-statistic values. The empty circle on the graph represents the F-statistic value, when the solid circle is the largest F-statistic, indicating the most effective number of groups at distinguishing the features and variables. As Fig. 6 shows, the F-statistic associated with five groups is highest, which is 17.1960. Seven group with the F-statistic 17.1249 is the second-best choice.

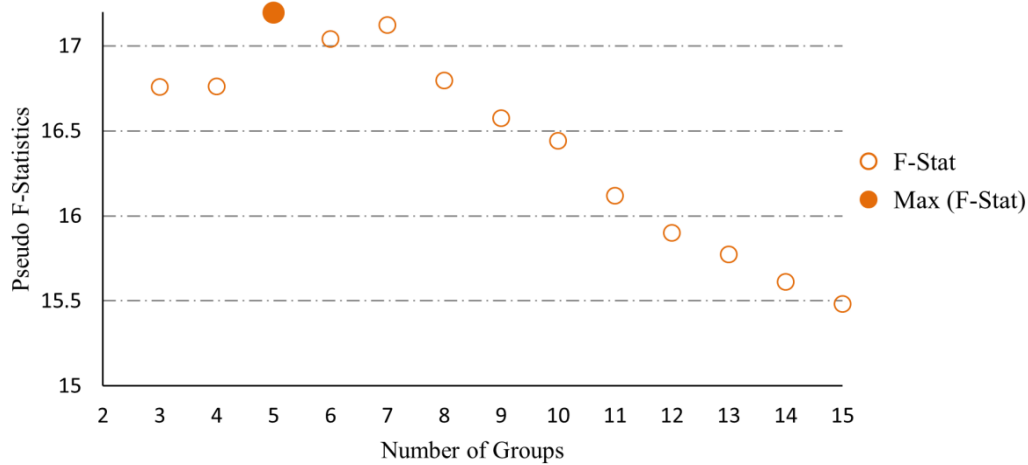

**Fig. 6.** Pseudo F-statistic plot in finding optimal number of groups

Therefor, all regions are classified into 5 clusters firstly, the result is shown in Fig. 7. The gray lines in the map illustrate the boundaries of provinces, when the white lines are the boundaries of prefecture-level cities in provinces.

**Cluster 1** is consists of the whole Shandong province and a small city from Henan province.

**Cluster 2** is consists of east China regions, include Shanghai city, Zhejiang province, Jiangsu province, Anhui province, Fujian province, Jiangxi province, part of Henan province.

**Cluster 3** includes the whole Hunan province, Guangdong province, Guangxi province and the east half part of Guizhou province.

**Cluster 4** is the North China, includes Beijing city, Tianjin city, Northeast regions, North west regions, Shanxi province, Hubei province, and the left half part of

Henan province, much of Qinghai province. It is worth mentioning that the separate area in cluster 7 is just an enclave of Haixi prefecture in Qinghai province.

**Cluster 5** is the southwest part of China. It includes the whole Sichuan province, Tibet Autonomous Region, Yushu prefecture from Qinghai province, Chongqing city (separated from Sichuan province in 1997), Yunnan province (except Wenshan prefecture), west half part of Guizhou province.

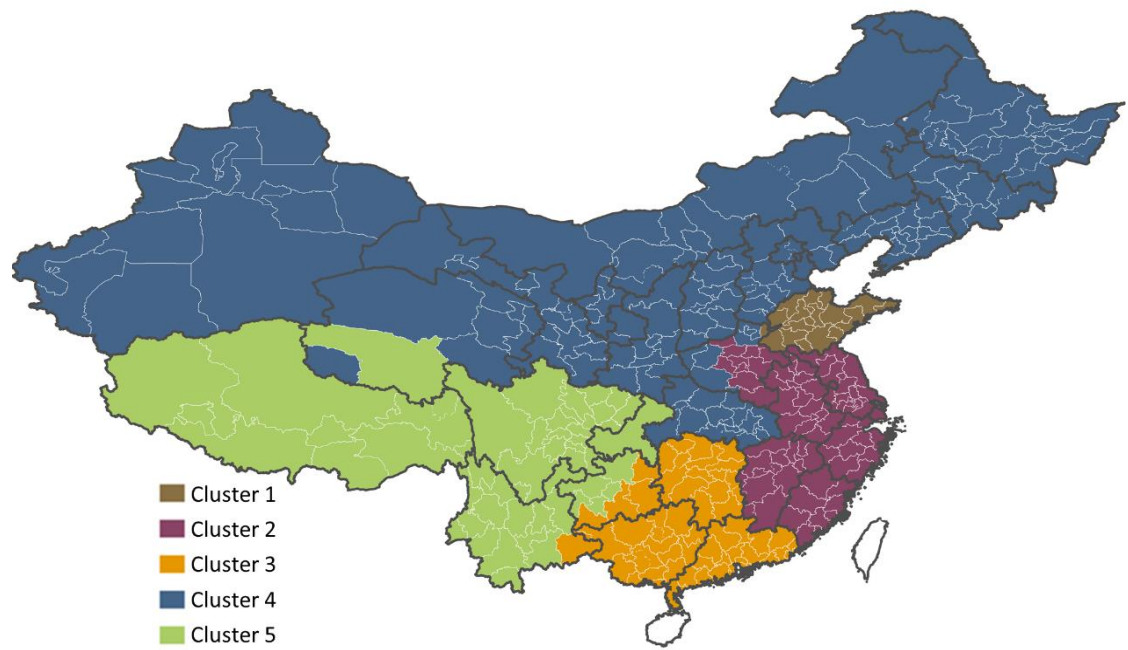

**Fig. 7.** Clustering of Chinese regions into 5 groups

If we choose the second-best choice and cluster the regions into 7 groups, as shown in Fig. 8, cluster 1, 2 and 5 don't have any change. Cluster 3 in Fig. 7 is splited into cluster 3 and 6 in Fig. 8, when cluster 4 in Fig. 7 is splited into cluster 4 and 7 in Fig. 8.

In Fig. 8, **Cluster 3** includes the whole Hunan province and Guilin city from Guangxi Autonomous Region.

**Cluster 4** is northeast region. It includes Heilongjiang province, Jilin province, Liaoning province (except Chaoyang city) and the north part of Inner Mongolia.

**Cluster 6** is mainly consists of Guangdong province, Guangxi province and the east half part of Guizhou province.

**Cluster 7** is the north and northwest regions of China, includes Beijing city, Tianjin city, Hebei province, Shanxi province, Shaanxi province, Hubei province, Gansu province, Ningxia Autonomous Region, Xinjiang Autonomous Region, much of Inner Mongolia Autonomous Region, left half part of Henan province, much of

Qinghai province. It is worth mentioning that the separate area in cluster 7 is just an enclave of Haixi prefecture in Qinghai province.

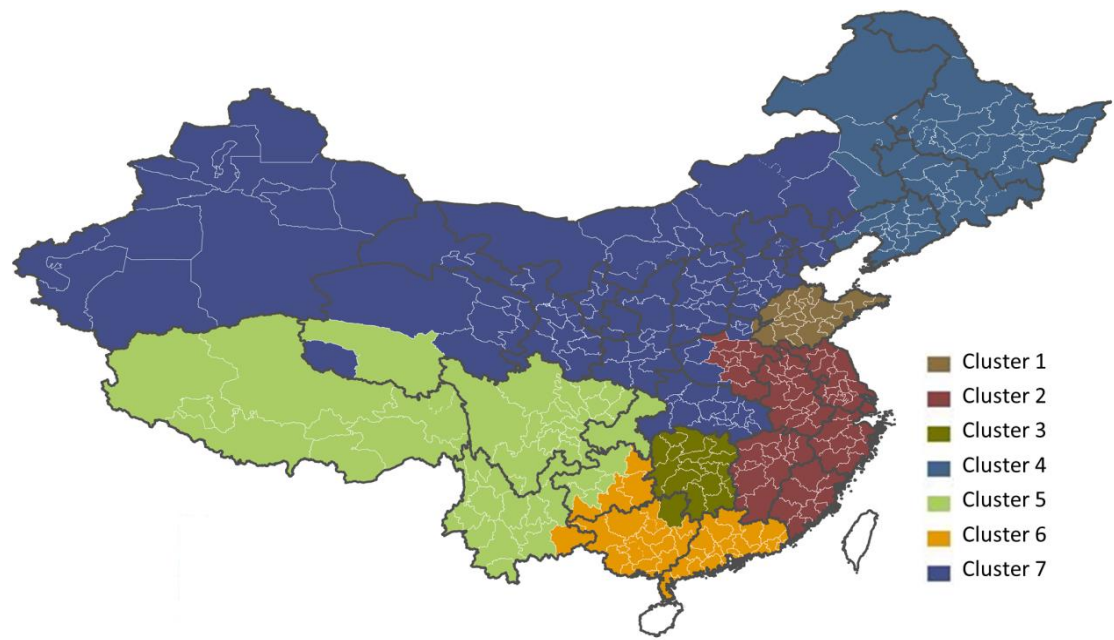

**Fig. 8.** Clustering of Chinese regions into 7 groups

# Discussion

Using geotagged online-gaming data of Chinese internet users, we cluster Chinese regions into several groups. As the clustering results show, the geographic boundaries of clusters coincide well with the boundaries of provincial regions. The north regions in China have more geographical proximity, when regional variations in south regions are more evident. Actually, the south part are the most populous and developed regions in China. However, unlike the wide flat plains in the north, south China is mountainous terrain, which has made the communication very difficult since ancient times.

# Acknowledgements

The work was supported by the project of "Social Science Foundation of China"[grant number 10CZX011); and the project of "Fundamental Research Funds for the Central Universities" (grant number DUT12RW309).

# References

Assunção, Renato Martins, Neves, Marcos Corrêa, Câmara, Gilberto, & Da Costa Freitas, C. (2006). Efficient regionalization techniques for socio‐economic geographical units using minimum spanning trees. *International Journal of Geographical Information Science, 20*(7), 797-811.

Brodbeck, Felix C, Frese, Michael, Akerblom, Staffan, Audia, Giuseppe, Bakacsi, Gyula, Bendova, Helena, . . . Brenk, Klas. (2000). Cultural variation of leadership prototypes across 22 European countries. *Journal of Occupational and Organizational Psychology, 73*(1), 1-29.

Caliński, Tadeusz, & Harabasz, Jerzy. (1974). A dendrite method for cluster analysis. *Communications in Statistics-theory and Methods, 3*(1), 1-27.

Cattell, Raymond B. (1950). The principal culture patterns discoverable in the syntal dimensions of existing nations. *The Journal of Social Psychology, 32*(2), 215-253.

Crang, M., & Zhang, J. (2012). Transient dwelling: trains as places of identification for the floating population of China. *Social & Cultural Geography, 13*(8), 895-914. doi: 10.1080/14649365.2012.728617

Eagle, Nathan, & Pentland, Alex. (2006). Reality mining: sensing complex social systems. *Personal Ubiquitous Computing, 10*(4), 255-268. doi: 10.1007/s00779-005-0046-3

Furnham, Adrian, Kirkcaldy, Bruce D, & Lynn, Richard. (1994). National attitudes to competitiveness, money, and work among young people: First, second, and third world differences. *Human Relations, 47*(1), 119-132.

Girardin, Fabien, Calabrese, Francesco, Fiore, Filippo Dal, Ratti, Carlo, & Blat, Jose. (2008). Digital footprinting: Uncovering tourists with user-generated content. *Pervasive Computing, IEEE, 7*(4), 36-43.

Gupta, Vipin, Hanges, Paul J, & Dorfman, Peter. (2002). Cultural clusters: methodology and findings. *Journal of world business, 37*(1), 11-15.

Haandrikman, K., & Hutter, I. (2012). That's a Different Kind of Person' - Spatial Connotations and Partner Choice. *Population Space and Place, 18*(3), 241-259. doi: 10.1002/psp.661

Ho, E. L. E., & Hatfield, M. E. (2011). Migration and Everyday Matters: Sociality and Materiality. *Population Space and Place, 17*(6), 707-713. doi: 10.1002/psp.636

Latour, Bruno. (2007). Beware, your imagination leaves digital traces. *Times Higher Literary Supplement, 6*(4), 2007.

Lazer, David, Pentland, Alex, Adamic, Lada, Aral, Sinan, Barabási, Albert-László, Brewer, Devon, . . . Van Alstyne, Marshall. (2009). Computational Social Science. *Science, 323*(5915), 721-723. doi: 10.1126/science.1167742

Manovich, Lev. (2011). Trending: the promises and the challenges of big social data. *Debates in the digital humanities*, 460-475.

O'Neill, Eamonn, Kostakos, Vassilis, Kindberg, Tim, Penn, Alan, Fraser, Danaë Stanton, & Jones, Tim. (2006). Instrumenting the city: Developing methods for observing and understanding the digital cityscape *UbiComp 2006: Ubiquitous Computing* (pp. 315-332): Springer.

Smith, Peter B, Dugan, Shaun, & Trompenaars, Fons. (1996). National culture and the values of organizational employees a dimensional analysis across 43 nations. *Journal of cross-cultural psychology, 27*(2), 231-264.

Wang, Xianwen, Peng, Lian, Zhang, Chunbo, Xu, Shenmeng, Wang, Zhi, Wang, Chuanli, & Wang, Xianbing. (2013). Exploring scientists' working timetable: A global survey. *Journal of*

*Informetrics, 7*(3), 665-675.

Wang, Xianwen, Xu, Shenmeng, Peng, Lian, Wang, Zhi, Wang, Chuanli, Zhang, Chunbo, & Wang, Xianbing. (2012). Exploring scientists' working timetable: Do scientists often work overtime? *Journal of Informetrics, 6*(4), 655-660.